\documentclass[twocolumn,amsmath,amssymb,floatfix,prl,aps,superscriptaddress,footinbib]{revtex4-1}

\usepackage{graphicx}
\usepackage{amsmath}
\usepackage{amssymb}
\usepackage{upgreek}
\usepackage{scrextend}
\usepackage{footmisc}

\def\micron{{\ \mu\text{m}}}                 % microns
\def\Hz{{\ \text{Hz}}}                       % Hz
\def\Rb87{^{87}\text{Rb}}                     % Rb 87
\def\Na23{^{23}\text{Na}}                     % Na 23
\def\Li6{^{6}\text{Li}}                       % Li 6
 \newcommand{\ket}[1]{\left|#1\right\rangle}

\DeclareMathAlphabet\mathbfcal{OMS}{cmsy}{b}{n}

\newcommand{\epst}{\tilde\epsilon}
\newcommand{\omx}{\tilde\omega_x}
\newcommand{\omy}{\tilde\omega_y}
\newcommand{\omz}{\tilde\omega_z}

\newcommand{\ex}{\mathbf{e}_x}
\newcommand{\ey}{\mathbf{e}_y}
\newcommand{\ez}{\mathbf{e}_z}
\newcommand{\benum}{\begin{enumerate}}
\newcommand{\eenum}{\end{enumerate}}
\newcommand{\OmC}{\Omega_{\rm C}}

\newcommand{\br}{\mathbf{r}}

\begin{document}

\title{Gauge matters: Observing the vortex-nucleation transition in a Bose condensate}

\author{L.~J.~LeBlanc}
\email[Correspondence to: ]{lindsay.leblanc@ualberta.ca}
\affiliation{Department of Physics, University of Alberta, Edmonton, AB T6G 2E1, Canada}
\affiliation{Joint Quantum Institute, National Institute of Standards and Technology, and University of Maryland, Gaithersburg, Maryland, 20899, USA}
\author{K.~Jim{\'e}nez-Garc{\'i}a}
\affiliation{Joint Quantum Institute, National Institute of Standards and Technology, and University of Maryland, Gaithersburg, Maryland, 20899, USA}
\affiliation{Departamento de F\'{\i}sica, Centro de Investigaci\'{o}n y Estudios Avanzados del Instituto Polit\'{e}cnico Nacional, M\'{e}xico D.F., 07360, M\'{e}xico}
\author{ R.~A.~Williams}
\altaffiliation[Current address: ]{National Physical Laboratory, Teddington TW11 0LW, UK}
\author{ M.~C.~Beeler}
\altaffiliation[Current address: ]{The Johns Hopkins Applied Physics Laboratory, Laurel, MD 20723, USA}
\author{W.~D.~Phillips}
\affiliation{Joint Quantum Institute, National Institute of Standards and Technology, and University of Maryland, Gaithersburg, Maryland, 20899, USA}
\author{I.~B.~Spielman}
\email[Correspondence to: ]{ian.spielman@nist.gov}
\affiliation{Joint Quantum Institute, National Institute of Standards and Technology, and University of Maryland, Gaithersburg, Maryland, 20899, USA}

\begin{abstract}
The order parameter of a quantum-coherent many-body system can include a phase degree of freedom, 
which, in the presence of an electromagnetic field, depends on the choice of gauge.   Because of the relationship between the phase gradient and the velocity, time-of-flight measurements reveal this gradient.
 Here, we make such measurements using initially trapped Bose-Einstein condensates (BECs) subject to an artificial magnetic field.  Vortices are nucleated in the BEC for artificial field strengths above a critical value, which represents a structural phase transition.   By comparing to superfluid-hydrodynamic  and Gross-Pitaevskii calculations,  we confirmed that the transition from the vortex-free state gives rise to a shear in the released BEC's spatial distribution, {representing a macroscopic method to measure this transition, distinct from direct measurements of vortex entry.} Shear is also affected by an artificial electric field accompanying the  artificial magnetic field turn-off, which depends on the details of the physical mechanism creating the artificial  fields,  and implies a natural choice of gauge. 
Measurements of this kind offer opportunities for studying  phase in less-well-understood quantum gas systems.  
\end{abstract}

\maketitle

While gauge invariance is central to our description of nature, specific physical situations lend themselves to a natural choice of gauge that matters.
A Bose-Einstein condensate's (BEC's) order parameter includes a gauge-dependent phase.
When a BEC is  subjected to sufficiently rapid rotation (or a sufficiently strong artificial magnetic field~\footnote{Since there is no real electric charge in this system, we work in terms of charge-free quantities, e.g., for the magnetic field $q\mathbf{B} \rightarrow \mathbfcal{B}$ and vector potential $q{\bf A}\rightarrow\mathbfcal{A}$.  Unless otherwise stated, the phrase ``magnetic field'' refers to a synthetic magnetic field.} $\mathcal{B}$),  the BEC exhibits vortices: points at which the density vanishes and the phase is  singular.  For a finite system, the structural phase transition from a state in which the phase varies smoothly to one with a single phase singularity occurs at a critical magnetic field $\mathcal{B}_{\rm cr}$, dependent on both particle-particle interactions and geometry~\cite{Fetter:2009fh}.  In quantum gas experiments, images of time-of-flight (TOF) expanded clouds can reveal the vortex-nucleation transition~\cite{Madison2000,Abo-Shaeer2001,Schweikhard2004,Fetter:2009fh,Lin2009b} via the appearance of local minima in the imaged atomic density, each associated with a  vortex core. Here, we detected this structural phase transition  via an abrupt shape-change in TOF-expanded BECs accompanying the appearance of vortices.  As we see below, this change depends upon the ``natural'' gauge choice for our experiment.

\begin{figure*}
\begin{center}
\includegraphics{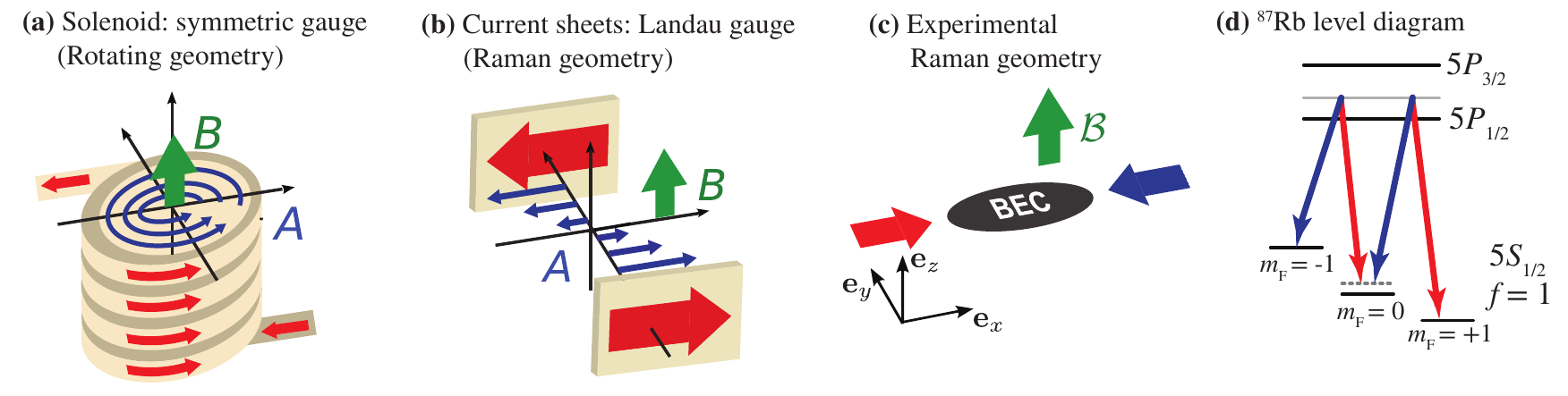}
\end{center}
\caption{(a) Solenoid  and (b) counterflowing current-sheet geometries, shown with  electrical current flow (red), vector potential $ \mathbf{A}$ (blue), and magnetic field $\mathbf{B}$ (green).   For each case, the natural gauge is illustrated: symmetric (suitable for rotating systems) $\mathbf{A}= (-{B}y\ex+{B}x\ey)/2$ in (a) and Landau (suitable for our Raman system) $\mathbf{A}= -{B}y\ex$ in (b).  (c) Experimental geometry. The elongated $^{87}$Rb BEC (black) experienced an artificial field $\mathbfcal{B}\propto\ez$  created using counterpropagating Raman lasers (red and blue arrows).  (d)  Level diagram of the $f=1$ ground state and excited $5P$ states, showing two-photon Raman transitions. }
\label{fig:setup}
\end{figure*}

In the description of many-body systems subject to a (real or synthetic) uniform magnetic field, the vector potential is generally written in the gauge most convenient for the problem at hand. Although the symmetric and Landau gauges are common choices, there is no \emph{a priori} ``natural'' gauge.  However, looking beyond the simple specification that the magnetic field is uniform, the physical mechanism that creates the magnetic field often suggests a specific gauge choice: any magnetic field created by a single electric current has a natural gauge determined by the geometry in which the vector potential is proportional to the electrical current.  In this sense, the symmetric gauge is a natural gauge for an infinite cylindrical solenoid [Fig.~1(a)], and the Landau gauge is a natural choice for two parallel counterflowing sheets of current [Fig~1(b)].  Analogous to an infinite solenoid, the synthetic magnetic field for rotating systems is most naturally expressed in the symmetric gauge $\mathbfcal{A}= (-\mathcal{B}y\ex+\mathcal{B}x\ey)/2$ because changes to the rotation rate lead to changes of $\mathbfcal{A}$ having the form of the symmetric gauge.  As in the case for two sheet currents,  our experiment's~\cite{lin2009}  synthetic magnetic field $\mathbfcal{B}$ is derived from an engineered vector potential~\cite{lin2009} that most naturally takes the Landau form $\mathbfcal{A}= -\mathcal{B}y\ex$, because the coupling between laser fields along $\ex$ only affects the component of $\mathbfcal{A}$ along $\ex$ [Fig.~1(c)].  Two experiments starting with the same $\mathbfcal{B}$ and ending with zero magnetic field can manifest different physical outcomes without violating gauge invariance, since the electric field $\mathbfcal{E} = -d\mathbfcal{A}/d t$ associated with the change~\cite{Lin2011a} is determined by the physical mechanism that creates $\mathbfcal{B}$.  In any gauge, the vector potential change $\Delta\boldsymbol{\mathcal{A}}$ resulting from modifications to experimental parameters is, in this interpretation of the natural gauge, directly proportional to  the {natural} gauge.

In this work, we studied $^{87}$Rb BECs subject to uniform laser-induced synthetic magnetic fields.  The constituent bosons 
experienced a Lorentz force, just as would charged particles in a magnetic field.  This synthetic magnetic field was continuously tunable from $0$ to above $\mathcal{B}_{\rm cr}$.  BECs are characterized by the complex-valued order parameter $\psi(\br) = \sqrt{\rho(\br)} \exp\left[i\phi(\br)\right]$ with superfluid density $\rho(\br)$ and phase $\phi(\br)$. 
The Gross-Pitaevskii equation (GPE) gives the time dependence of $\psi$. The single-valuedness of the phase lends the superfluid system  one of its defining properties: irrotationality.  
The current density $\mathbf{J}(\br) = \mathbb{R}\mathrm{e} [\psi^*(\br) \hat{\mathbf{v}}\psi(\br)]$, where $\hat {\bf v} = [-i\hbar\boldsymbol\nabla - \mathbfcal{A}(\br)]/m$,  describes the flow of particles and  links the local velocity to its phase $\phi(\br)$  via  ${\bf v}(\br) = \left[\hbar{\boldsymbol\nabla}\phi(\br) - {\mathbfcal A}(\br)\right]/m$ for particles of mass $m$. 
Notice that irrotationality applies to the local per-particle canonical momentum ${\bf p}(\br) = \hbar{\boldsymbol\nabla}\phi(\br)$, i.e., $\nabla \times {\bf p} = 0$, but  neither to velocity nor mechanical momentum $\mathbf{p}_{\rm m}(\br) = m\mathbf{v}(\br) =  \mathbf{p}(\br) - \mathbfcal{A(\br)}$.
 
We prepared trapped, equilibrated~\footnote{\label{f:note1}As in previous work~\cite{Lin2009b},  the equilibration time for the BEC in the artificial field was sufficient to reach equilibrium vortex density, but too small for the formation of an Abrikosov lattice of the vortex cores.} BECs in one of three configurations: ({\it i})  $\mathcal{B} = 0$, a reference case described by standard superfluid hydrodynamics [Fig.~2(a,d)]; ({\it ii})  $0< \mathcal{B} < \mathcal{B}_{\rm cr}$, described by modified superfluid hydrodynamics [Fig.~2(b,e)]; and ({\it iii})  $\mathcal{B} > \mathcal{B}_{\rm cr}$, with vortices  [Fig.~2(c,f)].  For all cases, we initiated TOF by abruptly removing the confining potential $V(\br)$ and rapidly making the  vector potential $\mathbfcal{A} =0$.  Effectively~\footnote{In practice, the final vector potential is spatially uniform but non-zero; as explained in the Supplementary materials, this does not affect the shape of the TOF-expanded BEC.}, this rapid turn-off left $\phi(\br)$ unaltered and mapped the gauge-dependent canonical momentum $\mathbf{p}$ just before TOF  (at $t_{0^-}$) onto the gauge-independent mechanical momentum  just after TOF began (at $t_{0^+}$):  $ \mathbf{p}(t_{0^-}) =  \mathbf{p}_{\rm m}(t_{0^+})$. 

Measuring the shearing motion in TOF, which has  contributions from the initial velocity field $\mathbf{v}(\br)$ and  the position-dependent electric force, allows us to distinguish between cases  ({\it i}),  ({\it ii}), and  ({\it iii}).

Case ({\it i}): The $\mathcal{B}=0$ expansion of a repulsively interacting BEC released from a harmonic trap is well-studied: zero-point energy is typically negligible and  interactions between the atoms dominate.  In the ground state, the constant (\emph{in situ}) phase gives $\mathbf{p}_{\rm m}(t_{0^+})=0$.  The familiar inverted-parabola Thomas-Fermi profile of harmonically trapped BECs~\cite{Anonymous:iieQCcLZ} is preserved by the interaction-driven expansion during TOF~{\cite{Castin:1996wp}}.

Case ({\it ii}): 
Modest magnetic fields ($0< \mathcal{B} < \mathcal{B}_{\rm cr}$) alter the interaction-dominated TOF expansion,
as was demonstrated in vortex-free rotating BEC systems~\cite{Hechenblaikner:2002ff}.
The \emph{in situ} phase $\phi(\br)$ is everywhere well-defined and the canonical momentum $\mathbf{p}(\br)$ differs from the mechanical momentum $\mathbf{p}_{\rm m}(\br)$ due to the presence of the vector potential that gives $\mathbfcal{B}={\boldsymbol \nabla} \times \mathbfcal{A}(\br)$.  The $t = 0^-$ canonical momentum in the Landau gauge
\begin{align}
\mathbf{p}^{ii} &=  -\mathcal{B}\left(\frac{\tilde\epsilon+1}{2}\right)  \left(y\ex +  x\ey\right)
\label{eq:irrotationalmomentum}
\end{align}
also defines $\mathbf{p}_{\rm m}(t_{0^+})$, where $-1< \tilde\epsilon< 1$ is a trap- and cyclotron-frequency-dependent anisotropy parameter [see Eq.~(\ref{eq:epst}) and Ref.~\onlinecite{Recati2001}].  Here the anisotropy is large, so that $\tilde{\epsilon}$ is nearly $-1$; compared to previous measurements~\cite{lin2009} in a cylindrically symmeteric trap ($\tilde\epsilon = 0$), the canonical momentum components are small. (In the corresponding symmetric gauge expression $\tilde\epsilon+1$ is replaced by $\tilde\epsilon$.)

Case ({\it iii}): Vortices  significantly affect the TOF expansion.  At the location of each vortex, the phase is singular [Fig.~2(c)], with a $2\pi$-winding  around it.  As an example, a vortex centered at the origin 
of a cylindrically symmetric system contributes a phase $\phi(\br) = \tan^{-1}(y/x)$, and the $t =  0^-$ canonical momentum in the Landau gauge is approximately 
\begin{align}
\mathbf{p}^{iii}  = \frac{\hbar \left(-y\ex + x\ey\right)}{x^2+y^2}-\mathcal{B}\left(\frac{\tilde\epsilon+1}{2}\right)\left(y\ex +  x\ey\right).
\end{align}
As before, this in-trap canonical momentum becomes the mechanical momentum $\mathbf{p}_{\rm m}(t_{0^+})$ as TOF begins.

We estimate the relative shear of these configurations from their momenta at $t_{0^+}$. Case ({\it i}) is shear free. For case ({\it ii}), the momentum along $\ex$ has typical scale $ \left|\mathbf{p}_{\rm m}\cdot \ex\right| \approx (\tilde\epsilon+1)\mathcal{B} y/2$.  In the presence of many vortices [case ({\it iii}), in  the diffused vorticity limit~\cite{Cozzini2003}] this increases by $ \hbar N_{\rm v}y/R^2$, where $R$ is the characteristic system size and $N_{\rm v}$ is the total number of vortices.   The difference is proportional to  $\hbar n_{\rm v} y$, where $n_{\rm v}$ is the areal vortex density, leading to an abrupt increase in the shearing momentum when vortices enter [Fig.~(\ref{fig:setup}g-i)]. 
Microscopically, this increased velocity originates from the spatial variations in phase  associated with the vortices [Fig.~2(c)].

The transition between configurations ({\it ii}) and ({\it iii}) at $\mathcal{B}_{\rm cr}$ occurs when the system can lower its energy by admitting a vortex; this critical field depends on the system's trap and interaction parameters~\cite{Fetter:2009fh}.  The critical cyclotron frequency $\OmC^{\rm cr} = \mathcal{B}_{\rm cr}/m$, above which  vortices are energetically stable~\cite{Fetter:2009fh,Svidzinsky2000}  { may be estimated as} $\OmC^{\rm crit}=(5\hbar/mR_{\perp}^2) \ln(R_{\perp}/\xi)$.  $R_{\perp} = \left[4\mu/m(\omega_x^2 + \omega_y^2)\right]^{1/2}$is the mean transverse Thomas-Fermi radius in a harmonic potential with frequencies $\omega_{x,y,z}$.  The healing length $\xi = (\hbar^2/2m\mu)^{1/2}$ sets the characteristic vortex core size, where $\mu = n_0(4\pi\hbar^2 a_s /m)$ is the central mean-field energy, $a_{\rm s}$ is the s-wave scattering length~\cite{Widera:2006cr}, and $n_0$ is the central density.  For the parameters in this experiment, $\OmC^{\rm cr}/2\pi = 12.0(1.1)$~Hz.

\begin{figure}[t!]
\begin{center}
\includegraphics{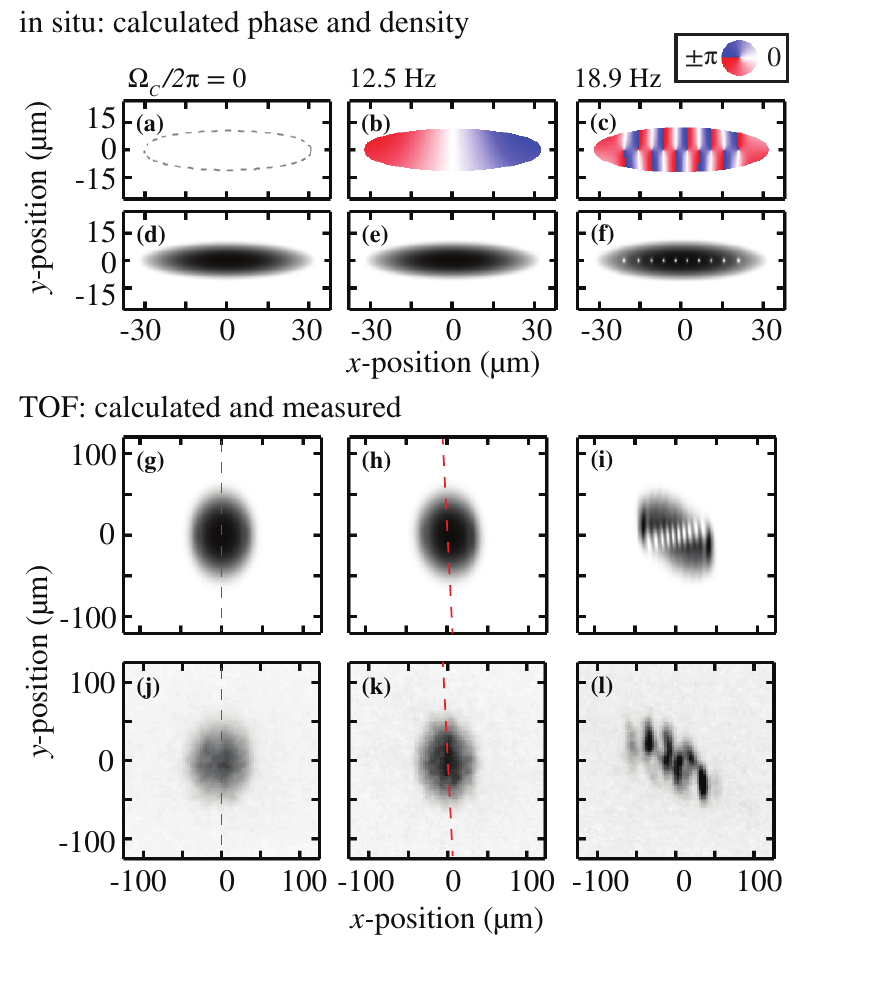}
\end{center}
\caption{GPE calculations of (a-f) {\emph{in situ}} and (g-i) time-of-flight distributions.  (a,b,c)  Landau-gauge-\emph{in-situ} phase,  (d,e,f) \emph{in situ} density, and (g,h,i) TOF density profiles in the $\ex$-$\ey$ plane for $\mathcal{B} < \mathcal{B}_{\rm cr}$  (left, centre) and $\mathcal{B} > \mathcal{B}_{\rm cr}$  (right).  The dashed grey ellipse in (a) indicates the Thomas-Fermi edge of the cloud; the phase is everywhere zero.  (j-l) Experimental TOF images for systems subject to magnetic fields with cyclotron frequencies, $\Omega_C = \mathcal{B}/m$.  A small degree of shear is evident in (k), but the largest shear is found in (l), where the cloud is also fragmented, due to the vortices' phase singularities. The dashed red lines in the lower images indicate the primary axis, to emphasize the small shear in (h) and (k).
}

\label{fig:setup}
\end{figure}

% ***********************************************************************************
% What we do
% ***********************************************************************************
We prepared BECs with $N \approx 1.4(3) \times10^5$  in the $f = 1$ ground state hyperfine manifold  at the intersection of two $\lambda = 1.064\micron$ laser beams [Fig.~1(c)].  The resulting potential was approximately harmonic and had measured frequencies $\{\omega_x, \omega_y, \omega_z\}/2\pi = \{10.1(1), 47.3(3), 90(1)\} \Hz$~\footnote{See attached Supplementary Materials for experimental and numerical details\label{f:supp}}. 
 We implemented an artificial magnetic field~\cite{Lin2009b,LeBlanc2012} using the combination of two counterpropagating $\lambda_{\rm R} = 790.1$~nm Raman lasers [Fig.~1(d)] traveling along $\pm\ex$ in conjunction with a (real) magnetic field $\mathbf{B} \cong(B_0+B^\prime y)\ey$, giving a gradient in detuning from Raman resonance $g_F \mu_{\rm B} B^\prime / h$ along $\ey$ ranging from 0 to $640\ {\rm Hz}/\mu{\rm m}$.  This gave effective cyclotron frequencies ranging from $\Omega_{\rm C}/2 \pi = 0$ to $20$~Hz. 

To study the evolution of the density distribution after mean-field-driven expansion, we released the atoms from the trap, and adiabatically transformed the
Raman-dressed superposition into a single Zeeman level for imaging~\cite{Williams:2012gs} in the first 2~ms of TOF~\footnotemark[\value{footnote}].  The cloud expanded for a total of 36.2~ms TOF before being imaged along $\ez$ [Fig.~\ref{fig:setup}(j,k,l)].

Figure~\ref{fig:setup}(a-f) shows computed \emph{in situ} phase and density distributions for a range of cyclotron frequencies.  Vortices nucleate only above $\Omega_{\rm cr}$, here only in panels (c) and (f).  For this geometry, the system's ground state consists of a linear chain of vortices; larger cyclotron frequencies or less anisotropy would result in a regular vortex lattice. 
(In the experiment, the vortices have not equilibrated to their ground state configuration~\footnotemark[2].)
Figure~\ref{fig:setup}(g-l) depicts calculated and measured TOF densities; the cloud's shear increases monotonically with increasing cyclotron frequency $\OmC = \mathcal{B}/m$ [e.g., there is a small  shear in Fig.~\ref{fig:setup}(h,k) and a large shear for Fig.~\ref{fig:setup}(i,l)].  
For strong artificial fields the observed density distributions were irregularly  fragmented [Fig.~\ref{fig:setup}(l)], while the computed clouds were ordered.  
These observations are both consistent with the presence of vortices -- disordered in the case of experiment -- whose characteristic phase gradients cause density modulations after TOF. 
The generic $\ey$-aligned stripes present in TOF [Fig.~\ref{fig:setup}(i,l)] result from the predominantly $\ey$-expansion from the anisotropic trap. 

\begin{figure}[tb!]
\begin{center}
\includegraphics{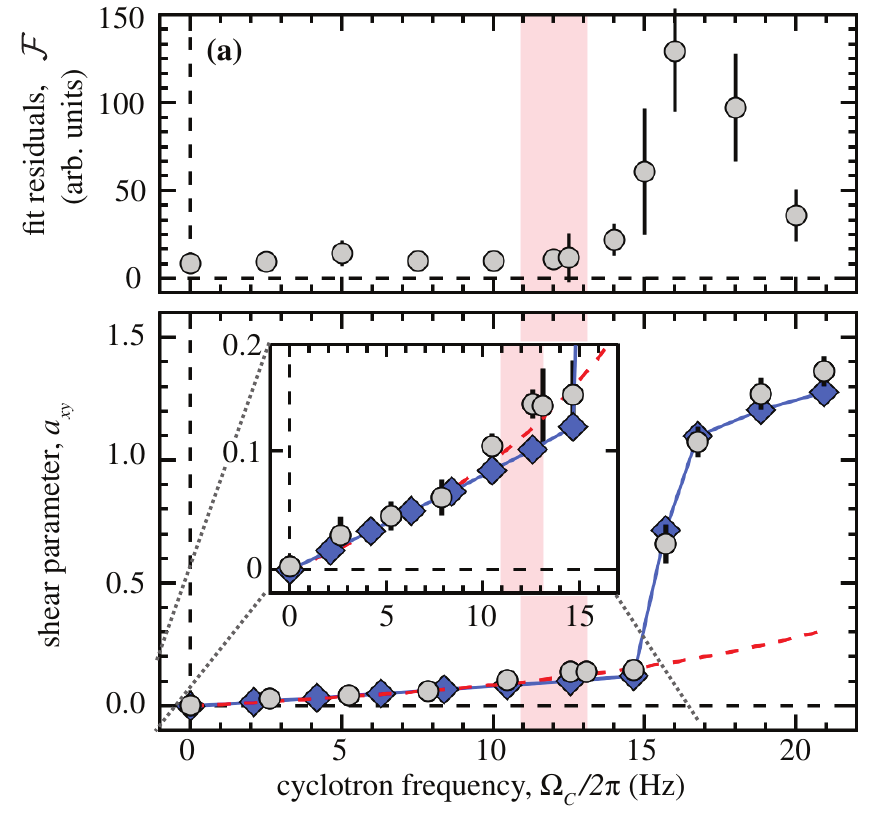}
\end{center}
\caption{
Signatures of vortex formation.  (a) Fit residuals, $\mathcal{F}$, indicating fragmentation due to vortices. (b) Shear  parameter $a_{\rm xy}$. The experimental data points (grey circles) each reflect an average over tens of measurements, and the uncertainty bars denote the  standard deviation of the mean [in (b) these are smaller than the symbol size].  Pink marks the  region, including uncertainty, where the vortex nucleation transition is expected: $\OmC/2\pi = 12.0(1.1)$~Hz. The shear parameter calculated for irrotational BECs is denoted by the red dashed curve and the results of the Raman GPE calculations (diamonds) connected by lines; neither has any free parameters. 
All uncertainties are statistical.}
\label{fig:results}
\end{figure}

\begin{figure}[tb!]
\begin{center}
\includegraphics{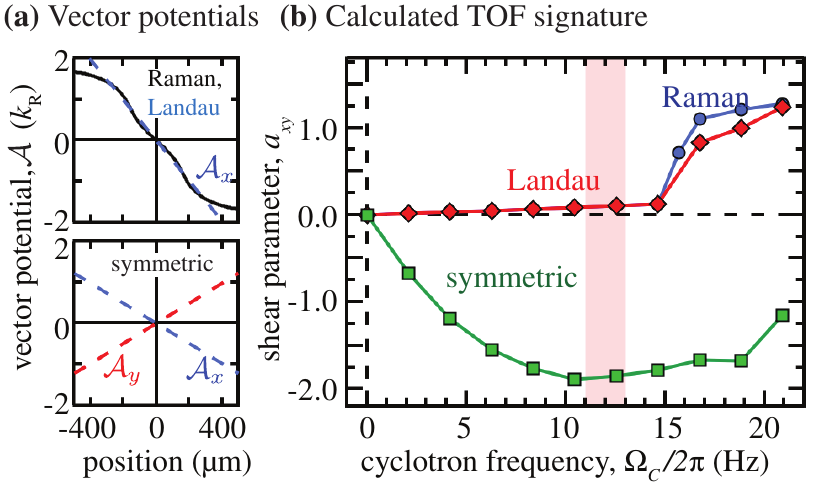}
\end{center}
\caption{
(a) Vector potentials used to calculate TOF response for cyclotron frequency $\Omega_C = 2\pi \times16.7$ Hz: the upper pannel shows the the vector potential corresponding to the exact Raman dispersion relationship (black) and the central-region fit to the Landau gauge (dashed blue) $\mathcal{A}_x= -\mathcal{B}y$ ($\mathcal{A}_y= 0$ in both cases), while the lower panel shows the symmetric gauge components $\mathcal{A}_x= -\mathcal{B}y/2$ (blue dashed) and $\mathcal{A}_y= \mathcal{B}x/2$ (red dashed). (b) Shear parameter $a_{xy}$ computed via GPE for equivalent-field systems after TOF and $\mathcal{B}$-turn-off for fields created by three physical mechanisms:  Raman (blue);   Landau gauge (red); and  symmetric gauge (green). 
}
\label{fig:gauges}
\end{figure}

We obtained TOF images at various artificial field strengths, and fit the resulting 2D column densities to a sheared  Thomas-Fermi profile 
\begin{align}
\label{eq:TFInt}
n^{\rm TOF}_{\rm2D} =  n_0^{\rm TOF}\left[1-\left(\frac{x}{R_x}\right)^2- \left(\frac{y}{R_y}\right)^2
+ a_{xy}\left(\frac{xy}{R_x R_y}\right)\right]^{3/2}
\end{align}
where $A$ is the $n_0^{\rm TOF}$ density; $R_{x,y}$ are Thomas-Fermi radii; and $a_{xy}$ is a dimensionless shear parameter. 
We obtain a measure of the fragmentation from the 
 fit residuals $\mathcal{F} = \sum_{x,y} [n_{\rm 2D}^{\rm meas}(x,y) - n_{\rm 2D}^{\rm fit}(x,y)]^2$.
Figure~\ref{fig:results}(a) shows that $\mathcal{F}$ sharply increases~\footnotemark[\value{footnote}] around $\OmC/2\pi = 15$~Hz, an indication of the transition to the BEC's vortex phase. 
 
Additional evidence for the entrance of vortices into this anisotropically trapped BEC is the behavior of the shear parameter $a_{xy}$:
Fig.~\ref{fig:results}(b) shows $a_{xy}$ sharply increasing above   $\OmC/2\pi = 15$~Hz,  in {reasonable} agreement with the predicted critical cyclotron frequency $\OmC^{\rm crit}/2\pi = 12.0(1.1)\Hz$.  We compare this result to two calculations. 

% Physicists: the atom's way of admiring itself.

The first calculation, the usual hydrodynamic description~\cite{Cozzini2003,Recati2001}, valid only in the absence of vortices, is modified to include the artificial magnetic field from a Landau-gauge vector potential~\cite{LeBlanc2012}, predicting the \emph{in situ} density
\begin{align}
\label{eq:TF}
&n(\br) =n_0\left[1-\left(\frac{x}{ \tilde R_x}\right)^2 - \left(\frac{y}{ \tilde R_y}\right)^2 - \left(\frac{z}{\tilde R_z}\right)^2\right]
\end{align}
and the canonical momentum $\mathbf{p}^{ii}$ [Eq.~\eqref{eq:irrotationalmomentum}].  Here, $\tilde R_i = [2\mu/m\tilde\omega_i^2]^{1/2}$ are  modified Thomas-Fermi radii with {effective trapping frequencies} $\tilde\omega_i$, where
$\omx^2 = \omega_x^2 + \Omega_C^2[(1+\epst)/2]^2$,  $\omy^2 = \omega_y^2 + \Omega_C^2[(1-\epst)/2]^2$,  and $\omz = \omega_z$.  The anisotropy parameter 
\begin{equation}
\epst = \frac{\omx^2 - \omy^2}{\omx^2 +\omy^2}
\label{eq:epst}
\end{equation}
can be obtained self-consistently from these equations~\cite{Cozzini2003,LeBlanc2012}.  These distributions are propagated in TOF using the hydrodynamic equations for comparison to the measured distributions.  From these calculated TOF distributions, we extracted the shear parameter and compared it with the measurement, as shown in Fig.~\ref{fig:results}(b).

The second calculation uses the GPE, which is valid for all our cases. We numerically found the GPE ground state at each $\mathcal{B}$ and then used the time-dependent GPE to calculate TOF evolution~\footnotemark[\value{footnote}].  
We find { excellent} agreement both with the amount of shear in the cloud for all configurations and in the location of the critical artificial field strength for vortex nucleation.

To understand the impact of the experimentally dictated natural gauge, we investigated the effects of different gauge choices.  
We performed GPE calculations as described above to obtain $a_{xy}$ for 
two alternate gauge choices with the same $\mathbfcal{B}$: the symmetric and Landau gauges.  
We compared these to the Raman system described above (which, as shown in Fig.~\ref{fig:gauges}(a), deviates only slightly from the ideal Landau gauge choice).
Though the Landau and symmetric gauge atomic systems are identical before release, their TOF 
 responses differ because of their different electric field impulses at turn-off, which reflect each system's natural gauge [Fig.~\ref{fig:gauges}(b)].   For $\mathcal{B} < \mathcal{B}_{\rm cr}$, this electric field impulse produces a significantly larger shearing when the symmetric gauge is natural.  This behavior reflects the fact that the gauge-dependent \emph{in situ} canonical momentum, which becomes   $\mathbf{p}_m(t_{0^+})$, is different in the two cases.
 Each system reacted to the entrance of vortices with a marked change in $a_{xy}$, here positive, as set by the vortices' direction of circulation, { but with a significantly different character.}
 
In conclusion, we showed that the shape of the cloud observed after TOF is strongly affected by the presence of vortices and signals the vortex nucleation transition.
By exploiting the connection between phase and velocity, we were able to extract topological features  of our system's order parameter from the TOF density distribution. Our experiment's anisotropic geometry and Landau-like natural gauge led to a signal that made this transition particularly clear.  { By comparing to what would have happened in the symmetric gauge, we showed that the gauge choice can make a significant difference in the experimental outcome.}
 In the future, techniques that match a system's geometry to the artificial field's natural gauge might be used to selectively excite edge modes in cold-atom quantum-Hall-like systems~\cite{Cazalilla:2005fp,Goldman:2013dg,Spielman:2013ck}.

\begin{acknowledgments}
We thank  J.~V.~Porto for useful conversations; S. Eckel, and D. G. Norris each for meticulously reading the manuscript; and J.~H.~Thywissen for the base code on which the GPE calculations were built.  This work was partially supported by the ONR; by the ARO with funds both from the DARPA-OLE  program and the Atomtronics MURI; and by the NSF through the Physics Frontier Center at JQI. L.J.L. acknowledges the NSERC of Canada, K.J.-G. acknowledges CONACYT, and M.C.B. acknowledges NIST-ARRA.  This research was undertaken, in part, thanks to funding from the Canada Research Chairs program.
\end{acknowledgments}

%%% ***   Set the bibliography file.   ***
%\bibliography{BibGeneral,BibPapers}

\begin{thebibliography}{27}%
\makeatletter
\providecommand \@ifxundefined [1]{%
 \@ifx{#1\undefined}
}%
\providecommand \@ifnum [1]{%
 \ifnum #1\expandafter \@firstoftwo
 \else \expandafter \@secondoftwo
 \fi
}%
\providecommand \@ifx [1]{%
 \ifx #1\expandafter \@firstoftwo
 \else \expandafter \@secondoftwo
 \fi
}%
\providecommand \natexlab [1]{#1}%
\providecommand \enquote  [1]{``#1''}%
\providecommand \bibnamefont  [1]{#1}%
\providecommand \bibfnamefont [1]{#1}%
\providecommand \citenamefont [1]{#1}%
\providecommand \href@noop [0]{\@secondoftwo}%
\providecommand \href [0]{\begingroup \@sanitize@url \@href}%
\providecommand \@href[1]{\@@startlink{#1}\@@href}%
\providecommand \@@href[1]{\endgroup#1\@@endlink}%
\providecommand \@sanitize@url [0]{\catcode `\\12\catcode `\$12\catcode
  `\&12\catcode `\#12\catcode `\^12\catcode `\_12\catcode `\%12\relax}%
\providecommand \@@startlink[1]{}%
\providecommand \@@endlink[0]{}%
\providecommand \url  [0]{\begingroup\@sanitize@url \@url }%
\providecommand \@url [1]{\endgroup\@href {#1}{\urlprefix }}%
\providecommand \urlprefix  [0]{URL }%
\providecommand \Eprint [0]{\href }%
\providecommand \doibase [0]{http://dx.doi.org/}%
\providecommand \selectlanguage [0]{\@gobble}%
\providecommand \bibinfo  [0]{\@secondoftwo}%
\providecommand \bibfield  [0]{\@secondoftwo}%
\providecommand \translation [1]{[#1]}%
\providecommand \BibitemOpen [0]{}%
\providecommand \bibitemStop [0]{}%
\providecommand \bibitemNoStop [0]{.\EOS\space}%
\providecommand \EOS [0]{\spacefactor3000\relax}%
\providecommand \BibitemShut  [1]{\csname bibitem#1\endcsname}%
\let\auto@bib@innerbib\@empty
%</preamble>
\bibitem [{Note1()}]{Note1}%
  \BibitemOpen
  \bibinfo {note} {Since there is no real electric charge in this system, we
  work in terms of charge-free quantities, e.g., for the magnetic field
  $q\protect \mathbf {B} \rightarrow \protect \mathbfcal {B}$ and vector
  potential $q{\protect \bf A}\rightarrow \protect \mathbfcal {A}$. Unless
  otherwise stated, the phrase ``magnetic field'' refers to a synthetic
  magnetic field.}\BibitemShut {Stop}%
\bibitem [{\citenamefont {Fetter}(2009)}]{Fetter:2009fh}%
  \BibitemOpen
  \bibfield  {author} {\bibinfo {author} {\bibfnamefont {A.}~\bibnamefont
  {Fetter}},\ }\href {\doibase 10.1103/RevModPhys.81.647} {\bibfield  {journal}
  {\bibinfo  {journal} {Rev. Mod. Phys.}\ }\textbf {\bibinfo {volume} {81}},\
  \bibinfo {pages} {647} (\bibinfo {year} {2009})}\BibitemShut {NoStop}%
\bibitem [{\citenamefont {Madison}\ \emph {et~al.}(2000)\citenamefont
  {Madison}, \citenamefont {Chevy}, \citenamefont {Wohlleben},\ and\
  \citenamefont {Dalibard}}]{Madison2000}%
  \BibitemOpen
  \bibfield  {author} {\bibinfo {author} {\bibfnamefont {K.~W.}\ \bibnamefont
  {Madison}}, \bibinfo {author} {\bibfnamefont {F.}~\bibnamefont {Chevy}},
  \bibinfo {author} {\bibfnamefont {W.}~\bibnamefont {Wohlleben}}, \ and\
  \bibinfo {author} {\bibfnamefont {J.}~\bibnamefont {Dalibard}},\ }\href
  {\doibase 10.1103/PhysRevLett.84.806} {\bibfield  {journal} {\bibinfo
  {journal} {Phys. Rev. Lett.}\ }\textbf {\bibinfo {volume} {84}},\ \bibinfo
  {pages} {806} (\bibinfo {year} {2000})}\BibitemShut {NoStop}%
\bibitem [{\citenamefont {Abo-Shaeer}\ \emph {et~al.}(2001)\citenamefont
  {Abo-Shaeer}, \citenamefont {Raman}, \citenamefont {Vogels},\ and\
  \citenamefont {Ketterle}}]{Abo-Shaeer2001}%
  \BibitemOpen
  \bibfield  {author} {\bibinfo {author} {\bibfnamefont {J.~R.}\ \bibnamefont
  {Abo-Shaeer}}, \bibinfo {author} {\bibfnamefont {C.}~\bibnamefont {Raman}},
  \bibinfo {author} {\bibfnamefont {J.~M.}\ \bibnamefont {Vogels}}, \ and\
  \bibinfo {author} {\bibfnamefont {W.}~\bibnamefont {Ketterle}},\ }\href
  {\doibase 10.1126/science.1060182} {\bibfield  {journal} {\bibinfo  {journal}
  {Science}\ }\textbf {\bibinfo {volume} {292}},\ \bibinfo {pages} {476}
  (\bibinfo {year} {2001})}\BibitemShut {NoStop}%
\bibitem [{\citenamefont {Schweikhard}\ \emph {et~al.}(2004)\citenamefont
  {Schweikhard}, \citenamefont {Coddington}, \citenamefont {Engels},
  \citenamefont {Mogendorff},\ and\ \citenamefont {Cornell}}]{Schweikhard2004}%
  \BibitemOpen
  \bibfield  {author} {\bibinfo {author} {\bibfnamefont {V.}~\bibnamefont
  {Schweikhard}}, \bibinfo {author} {\bibfnamefont {I.}~\bibnamefont
  {Coddington}}, \bibinfo {author} {\bibfnamefont {P.}~\bibnamefont {Engels}},
  \bibinfo {author} {\bibfnamefont {V.~P.}\ \bibnamefont {Mogendorff}}, \ and\
  \bibinfo {author} {\bibfnamefont {E.~A.}\ \bibnamefont {Cornell}},\ }\href
  {\doibase 10.1103/PhysRevLett.92.040404} {\bibfield  {journal} {\bibinfo
  {journal} {Phys. Rev. Lett.}\ }\textbf {\bibinfo {volume} {92}},\ \bibinfo
  {pages} {040404} (\bibinfo {year} {2004})}\BibitemShut {NoStop}%
\bibitem [{\citenamefont {Lin}\ \emph {et~al.}(2009{\natexlab{a}})\citenamefont
  {Lin}, \citenamefont {Compton}, \citenamefont {Jimenez-Garcia}, \citenamefont
  {Porto},\ and\ \citenamefont {Spielman}}]{Lin2009b}%
  \BibitemOpen
  \bibfield  {author} {\bibinfo {author} {\bibfnamefont {Y.-J.}\ \bibnamefont
  {Lin}}, \bibinfo {author} {\bibfnamefont {R.~L.}\ \bibnamefont {Compton}},
  \bibinfo {author} {\bibfnamefont {K.}~\bibnamefont {Jimenez-Garcia}},
  \bibinfo {author} {\bibfnamefont {J.~V.}\ \bibnamefont {Porto}}, \ and\
  \bibinfo {author} {\bibfnamefont {I.~B.}\ \bibnamefont {Spielman}},\
  }\href@noop {} {\bibfield  {journal} {\bibinfo  {journal} {Nature}\ }\textbf
  {\bibinfo {volume} {462}},\ \bibinfo {pages} {628} (\bibinfo {year}
  {2009}{\natexlab{a}})}\BibitemShut {NoStop}%
\bibitem [{\citenamefont {Lin}\ \emph {et~al.}(2009{\natexlab{b}})\citenamefont
  {Lin}, \citenamefont {Perry}, \citenamefont {Compton}, \citenamefont
  {Spielman},\ and\ \citenamefont {Porto}}]{lin2009}%
  \BibitemOpen
  \bibfield  {author} {\bibinfo {author} {\bibfnamefont {Y.-J.}\ \bibnamefont
  {Lin}}, \bibinfo {author} {\bibfnamefont {A.~R.}\ \bibnamefont {Perry}},
  \bibinfo {author} {\bibfnamefont {R.~L.}\ \bibnamefont {Compton}}, \bibinfo
  {author} {\bibfnamefont {I.~B.}\ \bibnamefont {Spielman}}, \ and\ \bibinfo
  {author} {\bibfnamefont {J.~V.}\ \bibnamefont {Porto}},\ }\href@noop {}
  {\bibfield  {journal} {\bibinfo  {journal} {Phys. Rev. A}\ }\textbf {\bibinfo
  {volume} {79}},\ \bibinfo {pages} {063631} (\bibinfo {year}
  {2009}{\natexlab{b}})}\BibitemShut {NoStop}%
\bibitem [{\citenamefont {Lin}\ \emph {et~al.}(2011)\citenamefont {Lin},
  \citenamefont {Compton}, \citenamefont {Jimenez-Garcia}, \citenamefont
  {Phillips}, \citenamefont {Porto},\ and\ \citenamefont
  {Spielman}}]{Lin2011a}%
  \BibitemOpen
  \bibfield  {author} {\bibinfo {author} {\bibfnamefont {Y.-J.}\ \bibnamefont
  {Lin}}, \bibinfo {author} {\bibfnamefont {R.~L.}\ \bibnamefont {Compton}},
  \bibinfo {author} {\bibfnamefont {K.}~\bibnamefont {Jimenez-Garcia}},
  \bibinfo {author} {\bibfnamefont {W.~D.}\ \bibnamefont {Phillips}}, \bibinfo
  {author} {\bibfnamefont {J.~V.}\ \bibnamefont {Porto}}, \ and\ \bibinfo
  {author} {\bibfnamefont {I.~B.}\ \bibnamefont {Spielman}},\ }\href@noop {}
  {\bibfield  {journal} {\bibinfo  {journal} {Nature Phys.}\ }\textbf {\bibinfo
  {volume} {7}},\ \bibinfo {pages} {531} (\bibinfo {year} {2011})}\BibitemShut
  {NoStop}%
\bibitem [{Note2()}]{Note2}%
  \BibitemOpen
  \bibinfo {note} {\label {f:note1}As in previous work~\cite {Lin2009b}, the
  equilibration time for the BEC in the artificial field was sufficient to
  reach equilibrium vortex density, but too small for the formation of an
  Abrikosov lattice of the vortex cores.}\BibitemShut {Stop}%
\bibitem [{Note3()}]{Note3}%
  \BibitemOpen
  \bibinfo {note} {In practice, the final vector potential is spatially uniform
  but non-zero; as explained in the Supplementary materials, this does not
  affect the shape of the TOF-expanded BEC.}\BibitemShut {Stop}%
\bibitem [{\citenamefont {Dalfovo}\ \emph {et~al.}(1999)\citenamefont
  {Dalfovo}, \citenamefont {Giorgini}, \citenamefont {Pitaevskii},\ and\
  \citenamefont {Stringari}}]{Anonymous:iieQCcLZ}%
  \BibitemOpen
  \bibfield  {author} {\bibinfo {author} {\bibfnamefont {F.}~\bibnamefont
  {Dalfovo}}, \bibinfo {author} {\bibfnamefont {S.}~\bibnamefont {Giorgini}},
  \bibinfo {author} {\bibfnamefont {L.~P.}\ \bibnamefont {Pitaevskii}}, \ and\
  \bibinfo {author} {\bibfnamefont {S.}~\bibnamefont {Stringari}},\ }\href@noop
  {} {\bibfield  {journal} {\bibinfo  {journal} {Rev. Mod. Phys.}\ }\textbf
  {\bibinfo {volume} {71}},\ \bibinfo {pages} {1} (\bibinfo {year}
  {1999})}\BibitemShut {NoStop}%
\bibitem [{\citenamefont {Castin}\ and\ \citenamefont
  {Dum}(1996)}]{Castin:1996wp}%
  \BibitemOpen
  \bibfield  {author} {\bibinfo {author} {\bibfnamefont {Y.}~\bibnamefont
  {Castin}}\ and\ \bibinfo {author} {\bibfnamefont {R.}~\bibnamefont {Dum}},\
  }\href@noop {} {\bibfield  {journal} {\bibinfo  {journal} {Phys. Rev. Lett.}\
  }\textbf {\bibinfo {volume} {77}},\ \bibinfo {pages} {5316} (\bibinfo {year}
  {1996})}\BibitemShut {NoStop}%
\bibitem [{\citenamefont {Hechenblaikner}\ \emph {et~al.}(2002)\citenamefont
  {Hechenblaikner}, \citenamefont {Hodby}, \citenamefont {Hopkins},
  \citenamefont {Marag{\`o}},\ and\ \citenamefont
  {Foot}}]{Hechenblaikner:2002ff}%
  \BibitemOpen
  \bibfield  {author} {\bibinfo {author} {\bibfnamefont {G.}~\bibnamefont
  {Hechenblaikner}}, \bibinfo {author} {\bibfnamefont {E.}~\bibnamefont
  {Hodby}}, \bibinfo {author} {\bibfnamefont {S.}~\bibnamefont {Hopkins}},
  \bibinfo {author} {\bibfnamefont {O.}~\bibnamefont {Marag{\`o}}}, \ and\
  \bibinfo {author} {\bibfnamefont {C.}~\bibnamefont {Foot}},\ }\href {\doibase
  10.1103/PhysRevLett.88.070406} {\bibfield  {journal} {\bibinfo  {journal}
  {Phys. Rev. Lett.}\ }\textbf {\bibinfo {volume} {88}},\ \bibinfo {pages}
  {070406} (\bibinfo {year} {2002})}\BibitemShut {NoStop}%
\bibitem [{\citenamefont {Recati}\ \emph {et~al.}(2001)\citenamefont {Recati},
  \citenamefont {Zambelli},\ and\ \citenamefont {Stringari}}]{Recati2001}%
  \BibitemOpen
  \bibfield  {author} {\bibinfo {author} {\bibfnamefont {A.}~\bibnamefont
  {Recati}}, \bibinfo {author} {\bibfnamefont {F.}~\bibnamefont {Zambelli}}, \
  and\ \bibinfo {author} {\bibfnamefont {S.}~\bibnamefont {Stringari}},\ }\href
  {\doibase 10.1103/PhysRevLett.86.377} {\bibfield  {journal} {\bibinfo
  {journal} {Phys. Rev. Lett.}\ }\textbf {\bibinfo {volume} {86}},\ \bibinfo
  {pages} {377} (\bibinfo {year} {2001})}\BibitemShut {NoStop}%
\bibitem [{\citenamefont {Cozzini}\ and\ \citenamefont
  {Stringari}(2003)}]{Cozzini2003}%
  \BibitemOpen
  \bibfield  {author} {\bibinfo {author} {\bibfnamefont {M.}~\bibnamefont
  {Cozzini}}\ and\ \bibinfo {author} {\bibfnamefont {S.}~\bibnamefont
  {Stringari}},\ }\href {\doibase 10.1103/PhysRevA.67.041602} {\bibfield
  {journal} {\bibinfo  {journal} {Phys. Rev. A}\ }\textbf {\bibinfo {volume}
  {67}},\ \bibinfo {pages} {041602} (\bibinfo {year} {2003})}\BibitemShut
  {NoStop}%
\bibitem [{\citenamefont {Svidzinsky}\ and\ \citenamefont
  {Fetter}(2000)}]{Svidzinsky2000}%
  \BibitemOpen
  \bibfield  {author} {\bibinfo {author} {\bibfnamefont {A.~A.}\ \bibnamefont
  {Svidzinsky}}\ and\ \bibinfo {author} {\bibfnamefont {A.~L.}\ \bibnamefont
  {Fetter}},\ }\href {\doibase 10.1103/PhysRevLett.84.5919} {\bibfield
  {journal} {\bibinfo  {journal} {Phys. Rev. Lett.}\ }\textbf {\bibinfo
  {volume} {84}},\ \bibinfo {pages} {5919} (\bibinfo {year}
  {2000})}\BibitemShut {NoStop}%
\bibitem [{\citenamefont {Widera}\ \emph {et~al.}(2006)\citenamefont {Widera},
  \citenamefont {Gerbier}, \citenamefont {F{\"o}lling}, \citenamefont
  {Gericke}, \citenamefont {Mandel},\ and\ \citenamefont
  {Bloch}}]{Widera:2006cr}%
  \BibitemOpen
  \bibfield  {author} {\bibinfo {author} {\bibfnamefont {A.}~\bibnamefont
  {Widera}}, \bibinfo {author} {\bibfnamefont {F.}~\bibnamefont {Gerbier}},
  \bibinfo {author} {\bibfnamefont {S.}~\bibnamefont {F{\"o}lling}}, \bibinfo
  {author} {\bibfnamefont {T.}~\bibnamefont {Gericke}}, \bibinfo {author}
  {\bibfnamefont {O.}~\bibnamefont {Mandel}}, \ and\ \bibinfo {author}
  {\bibfnamefont {I.}~\bibnamefont {Bloch}},\ }\href {\doibase
  10.1088/1367-2630/8/8/152} {\bibfield  {journal} {\bibinfo  {journal} {New J.
  Phys.}\ }\textbf {\bibinfo {volume} {8}},\ \bibinfo {pages} {152} (\bibinfo
  {year} {2006})}\BibitemShut {NoStop}%
\bibitem [{Note4()}]{Note4}%
  \BibitemOpen
  \bibinfo {note} {See attached Supplementary Materials for experimental and
  numerical details\label {f:supp}}\BibitemShut {NoStop}%
\bibitem [{\citenamefont {LeBlanc}\ \emph
  {et~al.}(2012{\natexlab{a}})\citenamefont {LeBlanc}, \citenamefont
  {Jimenez-Garcia}, \citenamefont {Williams}, \citenamefont {Beeler},
  \citenamefont {Perry}, \citenamefont {Phillips},\ and\ \citenamefont
  {Spielman}}]{LeBlanc2012}%
  \BibitemOpen
  \bibfield  {author} {\bibinfo {author} {\bibfnamefont {L.~J.}\ \bibnamefont
  {LeBlanc}}, \bibinfo {author} {\bibfnamefont {K.}~\bibnamefont
  {Jimenez-Garcia}}, \bibinfo {author} {\bibfnamefont {R.~A.}\ \bibnamefont
  {Williams}}, \bibinfo {author} {\bibfnamefont {M.~C.}\ \bibnamefont
  {Beeler}}, \bibinfo {author} {\bibfnamefont {A.~R.}\ \bibnamefont {Perry}},
  \bibinfo {author} {\bibfnamefont {W.~D.}\ \bibnamefont {Phillips}}, \ and\
  \bibinfo {author} {\bibfnamefont {I.~B.}\ \bibnamefont {Spielman}},\ }\href
  {\doibase 10.1073/pnas.1202579109} {\bibfield  {journal} {\bibinfo  {journal}
  {Proc. Natl. Acad. Sci. USA}\ }\textbf {\bibinfo {volume} {109}},\ \bibinfo
  {pages} {10811} (\bibinfo {year} {2012}{\natexlab{a}})}\BibitemShut {NoStop}%
\bibitem [{\citenamefont {Williams}\ \emph {et~al.}(2012)\citenamefont
  {Williams}, \citenamefont {LeBlanc}, \citenamefont {Jimenez-Garc{\'\i}a},
  \citenamefont {Beeler}, \citenamefont {Perry}, \citenamefont {Phillips},\
  and\ \citenamefont {Spielman}}]{Williams:2012gs}%
  \BibitemOpen
  \bibfield  {author} {\bibinfo {author} {\bibfnamefont {R.~A.}\ \bibnamefont
  {Williams}}, \bibinfo {author} {\bibfnamefont {L.~J.}\ \bibnamefont
  {LeBlanc}}, \bibinfo {author} {\bibfnamefont {K.}~\bibnamefont
  {Jimenez-Garc{\'\i}a}}, \bibinfo {author} {\bibfnamefont {M.~C.}\
  \bibnamefont {Beeler}}, \bibinfo {author} {\bibfnamefont {A.~R.}\
  \bibnamefont {Perry}}, \bibinfo {author} {\bibfnamefont {W.~D.}\ \bibnamefont
  {Phillips}}, \ and\ \bibinfo {author} {\bibfnamefont {I.~B.}\ \bibnamefont
  {Spielman}},\ }\href {\doibase 10.1126/science.1212652} {\bibfield  {journal}
  {\bibinfo  {journal} {Science}\ }\textbf {\bibinfo {volume} {335}},\ \bibinfo
  {pages} {314} (\bibinfo {year} {2012})}\BibitemShut {NoStop}%
\bibitem [{\citenamefont {Cazalilla}\ \emph {et~al.}(2005)\citenamefont
  {Cazalilla}, \citenamefont {Barberan},\ and\ \citenamefont
  {Cooper}}]{Cazalilla:2005fp}%
  \BibitemOpen
  \bibfield  {author} {\bibinfo {author} {\bibfnamefont {M.}~\bibnamefont
  {Cazalilla}}, \bibinfo {author} {\bibfnamefont {N.}~\bibnamefont {Barberan}},
  \ and\ \bibinfo {author} {\bibfnamefont {N.}~\bibnamefont {Cooper}},\ }\href
  {\doibase 10.1103/PhysRevB.71.121303} {\bibfield  {journal} {\bibinfo
  {journal} {Phys. Rev. B}\ }\textbf {\bibinfo {volume} {71}},\ \bibinfo
  {pages} {121303} (\bibinfo {year} {2005})}\BibitemShut {NoStop}%
\bibitem [{\citenamefont {Goldman}\ \emph {et~al.}(2013)\citenamefont
  {Goldman}, \citenamefont {Dalibard}, \citenamefont {Dauphin}, \citenamefont
  {Gerbier}, \citenamefont {Lewenstein}, \citenamefont {Zoller},\ and\
  \citenamefont {Spielman}}]{Goldman:2013dg}%
  \BibitemOpen
  \bibfield  {author} {\bibinfo {author} {\bibfnamefont {N.}~\bibnamefont
  {Goldman}}, \bibinfo {author} {\bibfnamefont {J.}~\bibnamefont {Dalibard}},
  \bibinfo {author} {\bibfnamefont {A.}~\bibnamefont {Dauphin}}, \bibinfo
  {author} {\bibfnamefont {F.}~\bibnamefont {Gerbier}}, \bibinfo {author}
  {\bibfnamefont {M.}~\bibnamefont {Lewenstein}}, \bibinfo {author}
  {\bibfnamefont {P.}~\bibnamefont {Zoller}}, \ and\ \bibinfo {author}
  {\bibfnamefont {I.~B.}\ \bibnamefont {Spielman}},\ }\href {\doibase
  10.1073/pnas.1300170110/-/DCSupplemental} {\bibfield  {journal} {\bibinfo
  {journal} {Proc. Nat. Acad. Sci. (USA)}\ }\textbf {\bibinfo {volume} {110}},\
  \bibinfo {pages} {6736} (\bibinfo {year} {2013})}\BibitemShut {NoStop}%
\bibitem [{\citenamefont {Spielman}(2013)}]{Spielman:2013ck}%
  \BibitemOpen
  \bibfield  {author} {\bibinfo {author} {\bibfnamefont {I.~B.}\ \bibnamefont
  {Spielman}},\ }\href {\doibase 10.1002/andp.201300110} {\bibfield  {journal}
  {\bibinfo  {journal} {Ann. Phys. (Berlin)}\ }\textbf {\bibinfo {volume}
  {525}},\ \bibinfo {pages} {797} (\bibinfo {year} {2013})}\BibitemShut
  {NoStop}%
%\bibitem [{\citenamefont {LeBlanc}\ \emph
%  {et~al.}(2012{\natexlab{b}})\citenamefont {LeBlanc}, \citenamefont
%  {Jim{\'e}nez-Garc{\'\i}a}, \citenamefont {Williams}, \citenamefont {Beeler},
%  \citenamefont {Perry}, \citenamefont {Phillips},\ and\ \citenamefont
%  {Spielman}}]{LeBlancPNAS2012}%
%  \BibitemOpen
%  \bibfield  {author} {\bibinfo {author} {\bibfnamefont {L.~J.}\ \bibnamefont
%  {LeBlanc}}, \bibinfo {author} {\bibfnamefont {K.}~\bibnamefont
%  {Jim{\'e}nez-Garc{\'\i}a}}, \bibinfo {author} {\bibfnamefont {R.~A.}\
%  \bibnamefont {Williams}}, \bibinfo {author} {\bibfnamefont {M.~C.}\
%  \bibnamefont {Beeler}}, \bibinfo {author} {\bibfnamefont {A.~R.}\
%  \bibnamefont {Perry}}, \bibinfo {author} {\bibfnamefont {W.~D.}\ \bibnamefont
%  {Phillips}}, \ and\ \bibinfo {author} {\bibfnamefont {I.~B.}\ \bibnamefont
%  {Spielman}},\ }\href {\doibase
%  10.1073/pnas.1202579109/-/DCSupplemental/pnas.1202579109_SI.pdf} {\bibfield
%  {journal} {\bibinfo  {journal} {Proc. Nat. Acad. Sci. (USA)}\ }\textbf
%  {\bibinfo {volume} {109}},\ \bibinfo {pages} {10811} (\bibinfo {year}
%  {2012}{\natexlab{b}})}\BibitemShut {NoStop}%
%\bibitem [{\citenamefont {Spielman}(2009)}]{Spielman2009}%
%  \BibitemOpen
%  \bibfield  {author} {\bibinfo {author} {\bibfnamefont {I.~B.}\ \bibnamefont
%  {Spielman}},\ }\href {\doibase 10.1103/PhysRevA.79.063613} {\bibfield
%  {journal} {\bibinfo  {journal} {Phys. Rev. A}\ }\textbf {\bibinfo {volume}
%  {79}},\ \bibinfo {eid} {063613} (\bibinfo {year} {2009})}\BibitemShut
%  {NoStop}%
%\bibitem [{\citenamefont {Dalfovo}\ and\ \citenamefont
%  {Stringari}(1996)}]{Dalfovo:1996uc}%
%  \BibitemOpen
%  \bibfield  {author} {\bibinfo {author} {\bibfnamefont {F.}~\bibnamefont
%  {Dalfovo}}\ and\ \bibinfo {author} {\bibfnamefont {S.}~\bibnamefont
%  {Stringari}},\ }\href
%  {http://eutils.ncbi.nlm.nih.gov/entrez/eutils/elink.fcgi?dbfrom=pubmed&id=9913160&retmode=ref&cmd=prlinks}
%  {\bibfield  {journal} {\bibinfo  {journal} {Phys. Rev. A}\ }\textbf {\bibinfo
%  {volume} {53}},\ \bibinfo {pages} {2477} (\bibinfo {year}
%  {1996})}\BibitemShut {NoStop}%
%\bibitem [{\citenamefont {Bao}\ \emph {et~al.}(2003)\citenamefont {Bao},
%  \citenamefont {Jin},\ and\ \citenamefont {Markowich}}]{Bao:2003gd}%
%  \BibitemOpen
%  \bibfield  {author} {\bibinfo {author} {\bibfnamefont {W.}~\bibnamefont
%  {Bao}}, \bibinfo {author} {\bibfnamefont {S.}~\bibnamefont {Jin}}, \ and\
%  \bibinfo {author} {\bibfnamefont {P.~A.}\ \bibnamefont {Markowich}},\ }\href
%  {\doibase 10.1137/S1064827501393253} {\bibfield  {journal} {\bibinfo
%  {journal} {SIAM J. Sci. Comput.}\ }\textbf {\bibinfo {volume} {25}},\
%  \bibinfo {pages} {27} (\bibinfo {year} {2003})}\BibitemShut {NoStop}%
\end{thebibliography}

\begin{thebibliography}{6}%
\makeatletter
%\providecommand \@ifxundefined [1]{%
% \@ifx{#1\undefined}
%}%
%\providecommand \@ifnum [1]{%
% \ifnum #1\expandafter \@firstoftwo
% \else \expandafter \@secondoftwo
% \fi
%}%
%\providecommand \@ifx [1]{%
% \ifx #1\expandafter \@firstoftwo
% \else \expandafter \@secondoftwo
% \fi
%}%
\providecommand \natexlab [1]{#1}%
\providecommand \enquote  [1]{``#1''}%
\providecommand \bibnamefont  [1]{#1}%
\providecommand \bibfnamefont [1]{#1}%
\providecommand \citenamefont [1]{#1}%
\providecommand \href@noop [0]{\@secondoftwo}%
\providecommand \href [0]{\begingroup \@sanitize@url \@href}%
\providecommand \@href[1]{\@@startlink{#1}\@@href}%
\providecommand \@@href[1]{\endgroup#1\@@endlink}%
\providecommand \@sanitize@url [0]{\catcode `\\12\catcode `\$12\catcode
  `\&12\catcode `\#12\catcode `\^12\catcode `\_12\catcode `\%12\relax}%
\providecommand \@@startlink[1]{}%
\providecommand \@@endlink[0]{}%
\providecommand \url  [0]{\begingroup\@sanitize@url \@url }%
\providecommand \@url [1]{\endgroup\@href {#1}{\urlprefix }}%
\providecommand \urlprefix  [0]{URL }%
\providecommand \Eprint [0]{\href }%
\providecommand \doibase [0]{http://dx.doi.org/}%
\providecommand \selectlanguage [0]{\@gobble}%
\providecommand \bibinfo  [0]{\@secondoftwo}%
\providecommand \bibfield  [0]{\@secondoftwo}%
\providecommand \translation [1]{[#1]}%
\providecommand \BibitemOpen [0]{}%
\providecommand \bibitemStop [0]{}%
\providecommand \bibitemNoStop [0]{.\EOS\space}%
\providecommand \EOS [0]{\spacefactor3000\relax}%
\providecommand \BibitemShut  [1]{\csname bibitem#1\endcsname}%
\let\auto@bib@innerbib\@empty
%</preamble>
\bibitem [{\citenamefont {LeBlanc}\ \emph {et~al.}(2012)\citenamefont
  {LeBlanc}, \citenamefont {Jim{\'e}nez-Garc{\'\i}a}, \citenamefont {Williams},
  \citenamefont {Beeler}, \citenamefont {Perry}, \citenamefont {Phillips},\
  and\ \citenamefont {Spielman}}]{LeBlancPNAS2012}%
  \BibitemOpen
  \bibfield  {author} {\bibinfo {author} {\bibfnamefont {L.~J.}\ \bibnamefont
  {LeBlanc}}, \bibinfo {author} {\bibfnamefont {K.}~\bibnamefont
  {Jim{\'e}nez-Garc{\'\i}a}}, \bibinfo {author} {\bibfnamefont {R.~A.}\
  \bibnamefont {Williams}}, \bibinfo {author} {\bibfnamefont {M.~C.}\
  \bibnamefont {Beeler}}, \bibinfo {author} {\bibfnamefont {A.~R.}\
  \bibnamefont {Perry}}, \bibinfo {author} {\bibfnamefont {W.~D.}\ \bibnamefont
  {Phillips}}, \ and\ \bibinfo {author} {\bibfnamefont {I.~B.}\ \bibnamefont
  {Spielman}},\ }\href {\doibase
  10.1073/pnas.1202579109/-/DCSupplemental/pnas.1202579109_SI.pdf} {\bibfield
  {journal} {\bibinfo  {journal} {Proc. Nat. Acad. Sci. (USA)}\ }\textbf
  {\bibinfo {volume} {109}},\ \bibinfo {pages} {10811} (\bibinfo {year}
  {2012})}\BibitemShut {NoStop}%
\bibitem [{\citenamefont {Lin}\ \emph {et~al.}(2011)\citenamefont {Lin},
  \citenamefont {Compton}, \citenamefont {Jimenez-Garcia}, \citenamefont
  {Phillips}, \citenamefont {Porto},\ and\ \citenamefont
  {Spielman}}]{Lin2011a}%
  \BibitemOpen
  \bibfield  {author} {\bibinfo {author} {\bibfnamefont {Y.-J.}\ \bibnamefont
  {Lin}}, \bibinfo {author} {\bibfnamefont {R.~L.}\ \bibnamefont {Compton}},
  \bibinfo {author} {\bibfnamefont {K.}~\bibnamefont {Jimenez-Garcia}},
  \bibinfo {author} {\bibfnamefont {W.~D.}\ \bibnamefont {Phillips}}, \bibinfo
  {author} {\bibfnamefont {J.~V.}\ \bibnamefont {Porto}}, \ and\ \bibinfo
  {author} {\bibfnamefont {I.~B.}\ \bibnamefont {Spielman}},\ }\href@noop {}
  {\bibfield  {journal} {\bibinfo  {journal} {Nature Phys.}\ }\textbf {\bibinfo
  {volume} {7}},\ \bibinfo {pages} {531} (\bibinfo {year} {2011})}\BibitemShut
  {NoStop}%
\bibitem [{\citenamefont {Spielman}(2009)}]{Spielman2009}%
  \BibitemOpen
  \bibfield  {author} {\bibinfo {author} {\bibfnamefont {I.~B.}\ \bibnamefont
  {Spielman}},\ }\href {\doibase 10.1103/PhysRevA.79.063613} {\bibfield
  {journal} {\bibinfo  {journal} {Phys. Rev. A}\ }\textbf {\bibinfo {volume}
  {79}},\ \bibinfo {eid} {063613} (\bibinfo {year} {2009})}\BibitemShut
  {NoStop}%
\bibitem [{\citenamefont {Dalfovo}\ and\ \citenamefont
  {Stringari}(1996)}]{Dalfovo:1996uc}%
  \BibitemOpen
  \bibfield  {author} {\bibinfo {author} {\bibfnamefont {F.}~\bibnamefont
  {Dalfovo}}\ and\ \bibinfo {author} {\bibfnamefont {S.}~\bibnamefont
  {Stringari}},\ }\href
  {http://eutils.ncbi.nlm.nih.gov/entrez/eutils/elink.fcgi?dbfrom=pubmed&id=9913160&retmode=ref&cmd=prlinks}
  {\bibfield  {journal} {\bibinfo  {journal} {Phys. Rev. A}\ }\textbf {\bibinfo
  {volume} {53}},\ \bibinfo {pages} {2477} (\bibinfo {year}
  {1996})}\BibitemShut {NoStop}%
\bibitem [{\citenamefont {Bao}\ \emph {et~al.}(2003)\citenamefont {Bao},
  \citenamefont {Jin},\ and\ \citenamefont {Markowich}}]{Bao:2003gd}%
  \BibitemOpen
  \bibfield  {author} {\bibinfo {author} {\bibfnamefont {W.}~\bibnamefont
  {Bao}}, \bibinfo {author} {\bibfnamefont {S.}~\bibnamefont {Jin}}, \ and\
  \bibinfo {author} {\bibfnamefont {P.~A.}\ \bibnamefont {Markowich}},\ }\href
  {\doibase 10.1137/S1064827501393253} {\bibfield  {journal} {\bibinfo
  {journal} {SIAM J. Sci. Comput.}\ }\textbf {\bibinfo {volume} {25}},\
  \bibinfo {pages} {27} (\bibinfo {year} {2003})}\BibitemShut {NoStop}%
\bibitem [{\citenamefont {Fetter}(2009)}]{Fetter:2009fh}%
  \BibitemOpen
  \bibfield  {author} {\bibinfo {author} {\bibfnamefont {A.}~\bibnamefont
  {Fetter}},\ }\href {\doibase 10.1103/RevModPhys.81.647} {\bibfield  {journal}
  {\bibinfo  {journal} {Rev. Mod. Phys.}\ }\textbf {\bibinfo {volume} {81}},\
  \bibinfo {pages} {647} (\bibinfo {year} {2009})}\BibitemShut {NoStop}%
\end{thebibliography}


\newpage
\appendix*
%\begin{widetext}
{\onecolumngrid
\begin{center}
{\bf\large Supplemental material for ``Gauge matters: Observing the BEC vortex-nucleation transition in a Bose condensate''} 

\vspace{12pt}
L.~J.~LeBlanc, K.~Jim{\'e}nez-Garc{\'i}a, R.~A.~Williams, M.~C.~Beeler, W.~D.~Phillips, I.~B.~Spielman
\end{center}
}
%\end{widetext}

\vspace{24pt}
\twocolumngrid

\renewcommand*{\citenumfont}[1]{S#1}
\renewcommand*{\bibnumfmt}[1]{[S#1]}
\renewcommand{\citenumfont}[1]{S#1}
\renewcommand{\thefigure}{S\arabic{figure}}
\setcounter{figure}{0}
\makeatletter
%\section{Experimental implementation}

\subsection{Experimental conditions}

{\bf Effects of modulation: }This manuscript and Ref.~\onlinecite{LeBlancPNAS2012} share the same underlying data set, put to very different purpose.  In this data set, the harmonic potential along $\bf{e_x}$ was periodically modulated, but sufficiently weakly that the system remained in the linear response regime. In the data presented here, release always occurred at the same phase in the modulation cycle. (Modulation was used to determine transport coefficients, which evidenced the superfluid Hall effect  reported in Ref.~\onlinecite{LeBlancPNAS2012}.)  For the present analysis, we chose those modulation times for which the shear is entirely due to the induced circulation due to the applied $\mathbfcal{B}$ and the vector-potential turn-off. We confirmed with our GPE simulations  that the modulation had no effect on the shapes of the clouds \emph{in situ} or in TOF. 

{\bf Effects of Raman dressing: } In addition to producing an artificial magnetic field, the combination of Raman dressing and real magnetic field gradient used in these experiments contributed  a scalar potential that locally depended on both Raman coupling strength and detuning from Raman resonance.  (The spatially-dependent, lowest-energy eigenstates of the system with Raman coupling plus magnetic field and field gradient along $\mathbf{e_y}$ experienced an effective harmonic antitrapping potential.)  This variation was measured experimentally by recording the frequency of dipole oscillation in the harmonic trap, confirming the weakened trapping potential along  $\mathbf{e_y}$.  We fit the resulting trap frequency to a phenomenological second-order polynomial as a function of cyclotron frequency and included this modification of the trapping potential in the superfluid hydrodynamic calculations: $\omega_y/2\pi = [c-a(\Omega_{\rm C}/2\pi+b)^2]$, where $a = 0.038(3)~({\rm Hz})^{-1}$ and $b = 2.1(9)~{\rm Hz}$ and $c = 47.5(4)$~{Hz} . We also fully accounted for the Raman-induced anistropic effective mass in this system~\cite{LeBlancPNAS2012,{Lin2011a}} in the superfluid hydrodynamics calculations.  In contrast, both the modified trapping potential and the effective mass are  automatically accounted for in the Raman-system GPE calculations, where the full dispersion relationship was used, and no  corrections are applied (see Notes on GPE calculations).

{\bf Atom number: } Our measurements of shear are relatively insensitive to atom-number differences between experimental realizations.  For $\mathcal{B} < \mathcal{B_{\rm cr}}$, the shear $a_{xy}$ is depends only weakly on $N$.  The atom number, $N = 1.4(3)\times10^5$ was determined from the TOF Thomas-Fermi radii $R_y$.  The uncertainty is the standard deviation of the measurements, and represents  shot-to-shot fluctuations.

\subsection{Trap and $\mathcal{B}$ turn-off}
The optical trapping potential was turned off at $t = 0$ in less than 1~$\mu$s. Concurrently, the artificial magnetic field was removed by adiabatically transforming the Raman-dressed superposition into a single Zeeman state, $\ket{f = 1, m_{F} = 1}$, by simultaneously ramping  the Raman intensity to zero and sweeping the bias magnetic field away from  Raman resonance.  This process was complete over the first 2~ms of TOF.  
Under this process, the 1/e time of turn-off of the vector potential was 130~$\mu$s and during this time, the cloud expanded only slightly while artificial field effects were significant.  Our GPE numerical calculations ignore the effects of this artificial field during the initial mean-field expansion and assume that all expansion occurred without the Raman beams present.  The agreement between our experimental and numerical results validate this assumption, as expected for an impulse that is short compared to typical timescales for expansion.   In the main text, we simplify the turn-off and consider  ``$t = 0^+$'' to be the time when all Raman coupling effects are removed.  

Due to the large final Raman detuning, the final vector potential $\mathbfcal{A}_{\rm f} = \mathcal{A}_{\rm f}\ex $ is non-zero, but uniform. This turn-off can be thought of as a two-step process. After  the first step, $\mathbfcal{A} \rightarrow 0$ and  the  mechanical momentum $\mathbf{p}^\prime_{\rm m}(t_{0+}) =\mathbf{p}_{\rm m}(t_{0-}) -\mathcal{B} y \ex$ [equal to the initial canonical momentum $\mathbf{p}(t_0)$] was changed by an amount proportional to $\mathbfcal{A}$ in the Landau gauge. Physically, this change in mechanical momentum resulted from the  electric field induced  as $\mathbfcal{A}\rightarrow0$.
 In the second step $\mathbfcal{A}= 0  \rightarrow \mathbfcal{A}= \mathbfcal{A}_f$, giving  the final mechanical momentum $\mathbf{p}_{\rm m}(t_{0^+}) = \mathbf{p}^\prime_{\rm m}(t_{0^+}) - \mathbfcal{A}_{\rm f}$, shifted by a constant $\mathbfcal{A}_f$ from the intermediate momentum $\mathbf{p}^\prime_{\rm m}(t_{0+})$, { because neither $\mathbfcal{A} = 0$ nor $\mathbfcal{A} = \mathbfcal{A}_f$ have any spatial variation.}  As it is the spatially dependence  that is of interest here, we consider in the main text the quantity  $\mathbf{p}^\prime_{\rm m}(t_{0+})$.  Effectively, the spatial variations of $\mathbf{p}^\prime_{\rm m}(t_{0+})$ and $\mathbf{p}_{\rm m}(t_{0+})$ are identical and independent of $\mathbfcal{A}_{\rm f}$, allowing us to consider only the $\mathbfcal{A}_{\rm f}=0$ case in the main text, where we use   $\mathbf{p}_{\rm m}(t_{0+})$.

\subsection{Notes on GPE calculations} 
We modelled our system using a 2+1 dimensional simulation, 
wherein we assumed a Thomas-Fermi profile along $\ez$, and solved the resulting 2D GPE in the $\ex$-$\ey$ plane, appropriately modified to account for the real 3D profile by including a position-dependent interaction factor~\cite{Spielman2009}.  We used imaginary-time propagation~\cite{Dalfovo:1996uc} to determine the initial wavefunctions, and evolved the system using a split-time spectral method~\cite{Bao:2003gd}. For the Raman-coupled GPE calculations, we assumed that the atoms remained in the lowest Raman dressed beam and used the energy versus momentum dispersion relation from the exact 3-level Raman-coupling Hamiltonian~\cite{Spielman2009} to describe the GPE's kinetic energy (which in this case is only modified along $\ex$).  In the presence of a detuning gradient along ${\bf e_y}$, this dispersion depended upon $y$.  This simulation accounts for both the non-uniformity of $\mathbfcal{B}$ and all contributions to the potential energy.
The calculation is valid for all three cases described in the main text, and for example correctly predicts the low-field shear,  $\Omega_{\rm C}$ for the onset of vortex nucleation, and the shear with vortices.

The regular structure in the simulated TOF distribution we see in case ({\it iii}) [Fig.~2(i)] results from the regularity of vortex positions \emph{in situ}.  If we start the GPE calculation with vortices seeded at random positions and allow the system to equilibrate, the vortices will eventually form a single row along the long axis, as seen in Fig.~2(f).  Figure S1 shows a simulation with the same parameters as  Fig.~2(i), but before equilibration.  This strongly resembles the experimental result with the same  parameters, Fig.~2(l).

\begin{figure}[tb!]
\begin{center}
\includegraphics{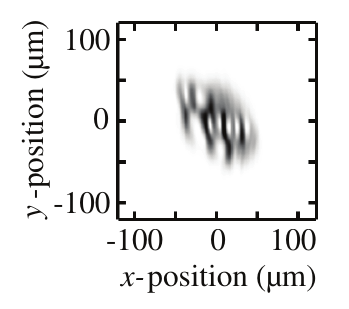}
\end{center}
\caption{GPE calculation of the time-of-flight distribution for a non-equilibrated vortex distribution with $\mathcal{B} > \mathcal{B}_{\rm cr}$; besides equilibration time, all parameters are identical to those in Fig.~2(i).  }
\label{fig:supp1}
\end{figure}

In our comparisons to the idealized gauge cases, we fit the vector potential computed for the Raman-coupled Hamiltonian near the trap centre to that of the Landau gauge for a uniform $\mathbfcal{B}$. We then used the same field value $\mathbfcal{B}$ for the symmetric gauge calculation. 
%The nonmonoticity of the symmetric gauge calculations in the high field regime is an artefact of the poor fits to the significantly fragmented distributions using sheared Thomas-Fermi fits when only a few vortices have entered the cloud.
 
 Note that the  cyclotron frequency differs by a factor of two  from the rotation frequency used in a similar expressions for rotating systems, as  in Ref.~\onlinecite{Fetter:2009fh} and elsewhere. 

 \vspace{20pt}
 \subsection{Vortex nucleation} 
 In the main text, we approximate the canonical momentum in case ({\it iii}) under the assumption of a single vortex at the centre of the cloud, and make the assumption that the healing length, i.e., the vortex core size,  is much smaller than the Thomas-Fermi radius. 

{\bf Fragmentation parameter:} The measured fragmentation parameter plotted in Fig.~3(a) in the regime of case ({\it iii}) is non-monotonic as a function of $\Omega_{\rm C}$ due to the details of the imaging process: though we see an initial increase as $\mathcal{B} $ increases beyond  $\mathcal{B}_{\rm cr}$, it does not continue to increase.  As more vortices enter the system, the TOF density variations increase in their spatial frequency, and due to the limited resolution of our imaging, this measure of fragmentation  decreases for higher spatial frequencies.

\end{document}